# Fabrication of (111)-oriented $Ca_{0.5}Sr_{0.5}IrO_3$/$SrTiO_3$ superlattices – a designed playground for honeycomb physics

Daigorou Hirai[1], Jobu Matsuno[2], and Hidenori Takagi[1, 3]

[1]*Department of Physics, University of Tokyo, Hongo 7-3-1, Bunkyo-ku, Tokyo 113-0033, Japan*

[2] *RIKEN Center for Emergent Matter Science (CEMS), Wako, Saitama 351-0198, Japan*

[3]*Max-Plank-Institute for solid state research, Heisenbergstrasse 1, Stuttgart 70569, Germany*

## Abstract

We report the fabrication of (111)-oriented superlattice structures with alternating $2m$-layers ($m$ = 1, 2, and 3) of $Ca_{0.5}Sr_{0.5}IrO_3$ perovskite and two layers of $SrTiO_3$ perovskite on $SrTiO_3$(111) substrates. In the case of $m$ = 1 bilayer films, the Ir sub-lattice is a buckled honeycomb, where a topological state may be anticipated. The successful growth of superlattice structures on an atomic level along the [111] direction was clearly demonstrated by superlattice reflections in x-ray diffraction patterns and by atomically-resolved transmission electron microscope images. The ground states of the superlattice films were found to be magnetic insulators, which may suggest the importance of electron correlations in Ir perovskites in addition to the much discussed topological effects.

## Introduction

The technical advances in the fabrication of oxide heterostructures on an atomic level have opened up a new avenue for exploring new materials by design, where the presence of interfaces may give rise to a rich variety of distinct electronic phases [1–4]. It was recently proposed that, by providing spatial constraint using heterostructure, a geometrical motif hidden in the underlying lattice can be isolated and generate topological phases [5–8]. If the two-layer units of perovskite-type transition-metal oxides (TMOs), $ABO_3$ ($B$ = transition metal), are isolated along the [111] crystallographic axis by forming a heterostructure with an insulating spacer, they can be viewed as a buckled honeycomb lattice (Fig. 1) of transition metal $B$ ions [5]. This bilayer of perovskite TMOs has a natively inverted band structure due to the geometry effect of the honeycomb lattice, which is similar to the situation in graphene [9]. With sizable spin-orbit coupling, a non-trivial band topology should be realized. (111)-oriented superlattices with bilayers of perovskite containing heavy $5d$ transition metals were therefore proposed to be promising candidates for topological matter.

Realization of (111)-oriented perovskite superlattices has been reported for only a few $3d$ systems because of the apparent technical difficulties [10–15]. Thin-film growth along [111] with atomic precision is challenging, because (111) surfaces are polar. SrTiO$_3$ (STO), for example, consists of alternate stacking of charged planes (SrO$_3^{4-}$ and Ti$^{4+}$) along the [111] direction. Its divergent surface energy makes it difficult to obtain a well-defined surface without complex surface reconstruction. In the case of $5d$ perovskites, due to the large radius of $5d$ ions, the lattice mismatch with commercially available $3d$ TMO substrates may bring another technical difficulty to film growth.

Here, we report the successful fabrication of (111)-oriented superlattices with $2m$ ($m$ = 1, 2, and 3) unit cells of $Ca_{0.5}Sr_{0.5}IrO_3$ and 2 unit cells $SrTiO_3$ [$(CSIO_{2m}, STO_2)_k$], grown epitaxially on STO(111) substrates by pulsed laser deposition. A solid-solution $Ca_{0.5}Sr_{0.5}IrO_3$ was used to stabilize (111)-oriented perovskite-type iridate films at the atomic level. Superlattice reflections observed in x-ray diffraction patterns and atomically-resolved transmission electron spectroscopy images indicate atomic ordering of $Ir^{4+}$ and $Ti^{4+}$ as designed.

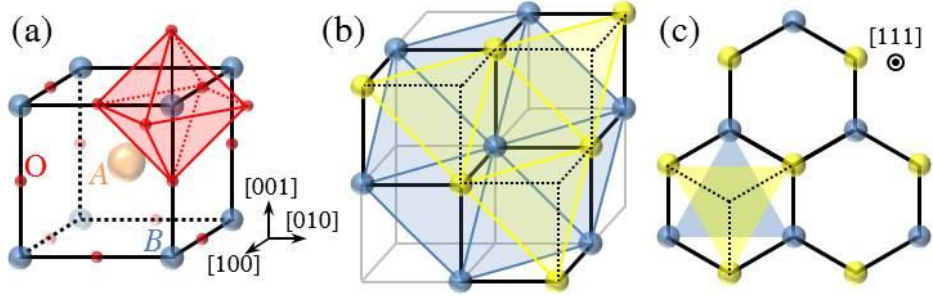

FIG. 1. (a) Schematic illustration of a cubic perovskite ($ABO_3$) unit cell. (b) (111) bilayer unit of $B$-sublattice with the two (111) planes indicated in yellow and blue. (c) The shortest $B$-$B$ bonds (solid line) in the bilayer projected on the (111) plane form a buckled honeycomb lattice.

### Experimental

Single-crystalline epitaxial thin films of $Ca_{0.5}Sr_{0.5}IrO_3$ (CSIO) were grown by pulsed laser deposition on surface treated STO(111) substrates. The details of the surface treatment, which produces an atomically flat surface, are given in Ref. [16]. Films were deposited using a KrF excimer laser ($\lambda$ = 248 nm) at 10 Hz with a fluency of ~ 1.5 J/cm$^2$. For the fabrication of (111)-oriented (CSIO$_{2m}$, STO$_2$)$_k$ superlattices, substrate temperatures and oxygen partial pressures were set at 720°C and 16 Pa during the deposition of CSIO and at 760°C and 1.5 Pa during the deposition of STO. The target used for the deposition of CSIO was a mixture of polycrystalline post-perovskite CaIrO$_3$ and monoclinically distorted six-layer type (6M) SrIrO$_3$, prepared by a conventional solid-state reaction. The quality and the lattice parameters of the films were characterized by x-ray diffraction (XRD) with Cu $K\alpha$ radiation on a diffractometer (Rigaku SmartLab) equipped with a Ge(002) monochromator. High-angle annular dark-field (HAADF) scanning transmission electron microscopy (STEM) images were taken using a STEM (JEOL JEM-ARM200F) at 200 kV at Foundation for Promotion of Material Science and Technology of Japan. The transport measurements were performed in a physical properties measurement system (PPMS, Quantum Design).

Results and discussion

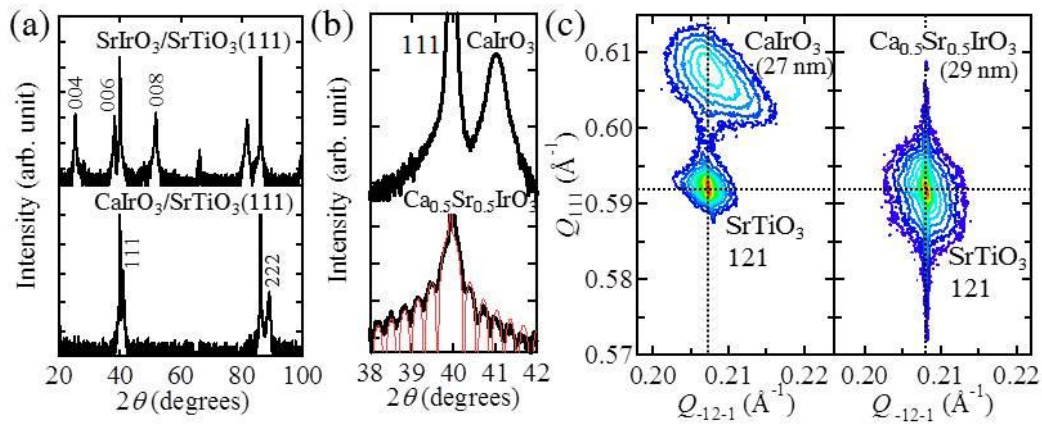

FIG. 2. (a) X-ray diffraction $\theta$-2$\theta$ scans of SrIrO$_3$ and CaIrO$_3$ films on SrTiO$_3$(111) substrate. (b) Magnified area of $\theta$-2$\theta$ scans to show the difference in the lattice

parameters and the thickness fringes between CaIrO$_3$ and Ca$_{0.5}$Sr$_{0.5}$IrO$_3$. The solid line represents a dynamical theory diffraction simulation. (c) Reciprocal-space maps around the 121 reflections for 27 nm thick CaIrO$_3$ and 29 nm thick Ca$_{0.5}$Sr$_{0.5}$IrO$_3$ films, where $Q_{\text{-}121}$ and $Q_{111}$ represent in-plane and out-of-plane reciprocal lattice vectors, respectively.

The (111)-oriented bilayer of perovskite SrIrO$_3$ has been discussed as a candidate for a topological insulator [5]. Under our experimental conditions, however, (111)-oriented perovskite SrIrO$_3$ was not stabilized on the STO(111) substrate. The $\theta$-$2\theta$ XRD scans for SrIrO$_3$ films deposited on STO(111) substrates [Fig. 2(a)] revealed that the (001)$_m$-oriented hexagonal-based 6M structure of SrIrO$_3$ [17] was formed, where the Ir$_2$O$_9$ units with two face-sharing octahedra are linked by the corners. The growth of the (001)$_m$-oriented 6M phase on STO(111) substrate by metal organic chemical vapor deposition [18] was previously reported. 6M SrIrO$_3$ seems to be more stable than perovskite SrIrO$_3$ on STO(111).

$AB$O$_3$ oxides have been known to exhibit a sequential transformation of crystal structure as a function of the size of the $A$-site cation or applied pressure. The evolution of the crystal structure can be described by the change in the stacking pattern of the two types of layer with corner-sharing and face-sharing $B$O$_6$ octahedra. In general, the ratio of number of corner (C) to face-sharing (F) layers in the stacking (C/F) increases as 2H(0) → 9R(1/2) → 6H(2/1) → perovskite($\infty$), with replacing smaller $A$-site cations or increasing pressure. This trend suggests that an isoelectronic compound CaIrO$_3$, with the smaller Ca$^{2+}$ ion rather than $Sr^{2+}$ on the *A*-site, could have a better chance of adopting a perovskite structure on STO(111) in contrast to the case for $SrIrO_3$.

The growth of perovskite $CaIrO_3$ films was attempted on STO(111) substrate. Perovskite $CaIrO_3$ was in fact stabilized, which is evidenced by the $\theta$-$2\theta$ XRD patterns, as shown in Fig. 2(a). Well-defined 111 and 222 peaks in the $\theta$-$2\theta$ XRD patterns indicate that a single crystalline (111)-oriented perovskite $CaIrO_3$ thin film was obtained without any impurity phases, such as post-perovskite $CaIrO_3$ or $IrO_2$. Reciprocal space mapping for the $CaIrO_3$ film on STO(111) around the 121 reflection, shown in Fig. 2(c), supports further the formation of epitaxially grown perovskite thin films. The in-plane lattice constant obtained from the 121 Bragg reflection agrees well with that of the STO substrate, implying that the majority of the $CaIrO_3$ film is coherently strained on the substrate. The anisotropic shape of the Bragg spot, however, suggests a partial strain relaxation at the film surface. The nominal lattice mismatch between perovskite $CaIrO_3$ ($a_{pc}$ ~ 3.855 Å; pseudo-cubic notation) and STO ($a$ = 3.905 Å) is relatively large (1.3%), which may result in a partial strain relaxation. Since the epitaxial strain can be released by a large surface deformation, the surface of strained films becomes unstable. This is known as the Asaro-Tiller-Grinfeld instability [19–21]. Indeed, no thickness fringe was seen around the 111 reflection of the $CaIrO_3$ film in the $\theta$-$2\theta$ XRD scan [Fig. 2(b)], which indicates that film thickness is not well-defined due to the disordered surface. Atomic force microscopy observations (not shown) support the hypothesis of significant surface roughness of the $CaIrO_3$ films grown on STO(111).

In order to obtain the atomically flat surface required to realize a superlattice structure, we attempted to reduce the lattice mismatch between iridium perovskite and the STO substrate by making a solid solution of $CaIrO_3$ and $SrIrO_3$. A solid solution of $CaIrO_3$ and $SrIrO_3$ can be formed only in a limited composition range by solid-state reaction [22]. In the case of epitaxial thin-film growth, however, the solid solution was reported to exist for any composition [23]. With partial Sr substitution for Ca, the unit cell volume of perovskite $Ca_{1-x}Sr_xIrO_3$ was increased. The best match of the lattice constant of $Ca_{1-x}Sr_xIrO_3$ with that of STO was achieved at around $x = 0.5$. As shown in Fig. 2(c), the position of the 121 Bragg spot of $Ca_{0.5}Sr_{0.5}IrO_3$ (CSIO) in the reciprocal space mapping almost perfectly overlaps with that of the STO substrate, indicating that both the in-plane and the out-of plane lattice constants of CSIO are equivalent to those of the STO substrate. By eliminating the lattice mismatch, the surface quality was drastically improved in CSIO films compared with those of pure $CaIrO_3$. The magnified view of the XRD pattern in Fig. 2(b) shows the clear contrast between pure $CaIrO_3$ and CSIO films. The thickness fringes arising from the finite thickness of the films were pronounced for CSIO, evidencing the drastically improved quality of the surface. The atomically flat surface of CSIO enabled us to construct superlattice structures.

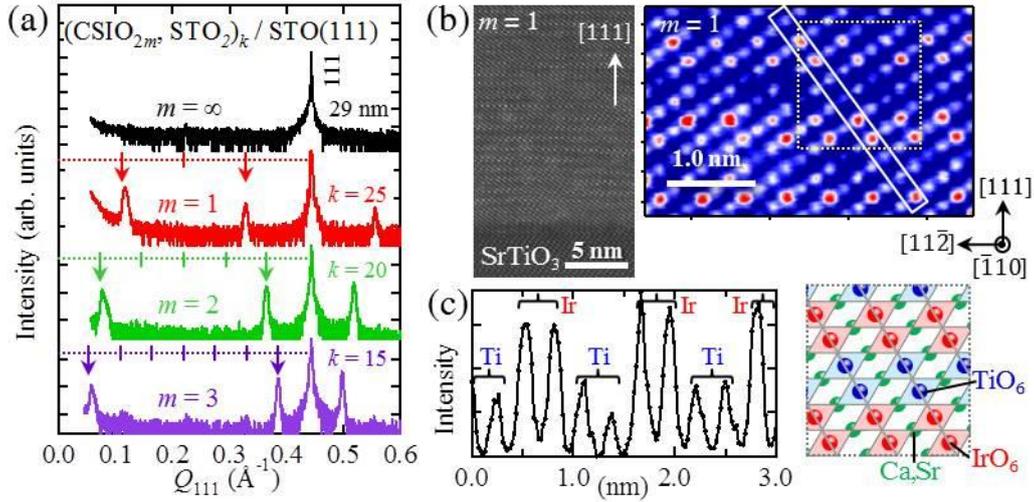

FIG. 3. (a) X-ray diffraction $\theta$-$2\theta$ scans for the superlattice [(Ca$_{0.5}$Sr$_{0.5}$IrO$_3$)$_{2m}$, (SrTiO$_3$)$_2$]$_k$ on SrTiO$_3$(111) with $m$ = 1, 2, 3, and $\infty$. Superlattice reflections are indicated by arrows. (b) Atomically-resolved HAADF-STEM image of the superlattice with $m$ = 1 along the SrTiO$_3$[111] direction. (c) Left panel; Intensity scan along the column indicated in the HAADF-STEM image shown in (b). Right panel; Schematic illustration of stacking of IrO$_6$ and TiO$_6$ (111) bilayers in the dotted square in (b).

By inserting insulating (111) bilayers of SrTiO$_3$ (Ti$^{4+}$, $d^0$) into the perovskite CSIO, (111) bilayer and other higher harmonics of bilayers of perovskite iridate were successfully isolated in the artificial superlattice. The superlattice films with alternating 2$m$ ($m$ = 1, 2, and 3) unit cells of perovskite Ca$_{0.5}$Sr$_{0.5}$IrO$_3$ and 2 unit cells of SrTiO$_3$ (CSIO$_{2m}$, STO$_2$) on STO(111) were grown with a thickness of 22 - 26 nm. The successful growth was confirmed by the $\theta$-$2\theta$ XRD scans and atomically-resolved HAADF-STEM observations, as shown in Fig. 3. Besides the fundamental peaks originating from the cubic perovskite lattice, clear satellite peaks were observed in the XRD patterns for all of the ($CSIO_{2m}$, $STO_2$) superlattices, indicating the growth of a superlattice and well-defined interfaces between CSIO and STO layers within given experimental resolution. The superlattice peaks indicate the periodicity of $(2m + 2)d$, where $d$ is the (111) interlayer distance of 2.25 Å, evidencing that the designed layering was achieved.

The ordering of $Ir^{4+}$ and $Ti^{4+}$ ions in the $m = 1$ (bilayer of Ir) superlattice is visualized by the [-110] zone axis HAADF-STEM image with strong atomic number (Z) contrast presented in Fig. 3(b). A stripe modulation of $Ir^{4+}$ (bright spots) and $Ti^{4+}$ (dark spots) represents an alternate stacking of CSIO and STO bilayers along the [111] direction. In the intensity scan in a column along the [001] direction, shown in Fig. 3(c), pairs of $Ir^{4+}$ (high intensity peaks) and $Ti^{4+}$ (low intensity peaks) ions emerge alternately as expected for a $m = 1$ superlattice. From these structural data, we conclude that the (111) bilayer of Ir, forming a buckled honeycomb lattice, and its higher harmonics were obtained in the coherently grown superlattice.

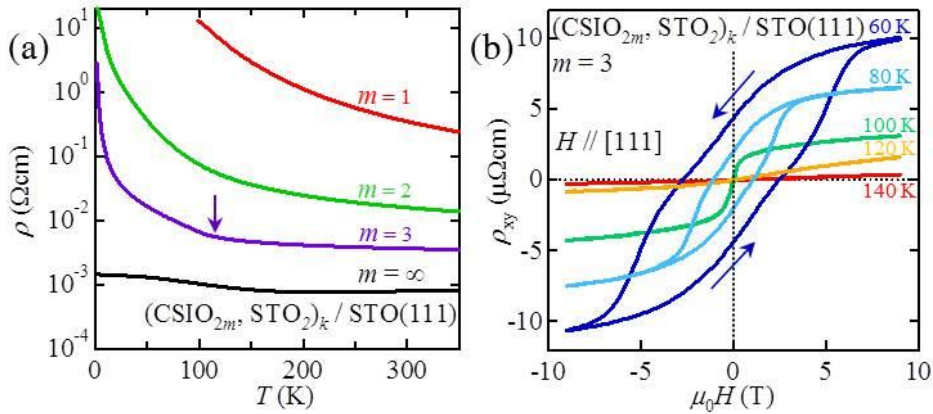

FIG. 4. (a) Temperature dependence of resistivity $\rho(T)$ for the superlattice [$(Ca_{0.5}Sr_{0.5}IrO_3)_{2m}$, $(SrTiO_3)_2$]$_k$ / $SrTiO_3$(111) with $m = 1, 2, 3$, and $\infty$. A kink in the resistivity curve for the $m = 3$ superlattice is indicated by an arrow. (b) Hall resistivity ($\rho_{xy}$) of the $m = 3$ superlattice at various temperatures is plotted as a function of applied field.

Resistivity $\rho(T)$ measurements were conducted on the grown (111) films. The results are summarized in Fig. 4(a), where metal-insulator transition as a function of number of CSIO bilayers can be seen. The CSIO only, $m = \infty$, (111) film showed a poorly metallic behavior of resistivity $\rho(T)$, almost temperature independent and with a magnitude ~ 1 mΩcm. This behavior is essentially the same as observed in bulk and thin-film samples of perovskite $CaIrO_3$ [23,24] and $SrIrO_3$ [25,26], consistent with a semimetallic ground state due to the presence of symmetry-protected Dirac nodes [27,28].

The insertion of (111) STO bilayers changes the ground state from semimetallic to insulating. All of the ($CSIO_{2m}$, $STO_2$) ($m = 1$, 2, and 3) superlattices showed an insulating behavior of $\rho(T)$ at low temperatures. With decreasing the number of CSIO (111) bilayers from $m = 3$ to 1, $\rho(T)$ systematically changed to more insulating behavior both in magnitude and temperature dependence. The (111)-oriented CSIO bilayer in the $m = 1$ superlattice was fully insulating below room temperature.

Furthermore, a metal to magnetic insulator transition was observed for the $m = 3$ superlattice. $\rho(T)$ of the $m = 3$ film showed a poorly metallic behavior similar to the $m = \infty$ film at high temperatures. On cooling, however, a clear kink in $\rho(T)$ followed by insulating behavior was observed at 100 K (indicated by an arrow). In the insulating state of the $m = 3$ superlattice below 100 K, an anomalous Hall effect was observed, as shown in Fig. 4(b). A clear hysteresis of Hall resistivity $\rho_{xy}(B)$ as a function of magnetic field perpendicular to the film plane was observed only below the kink temperature of 100 K, indicative of the presence of out-of-plane ferromagnetic moments in the insulating state. The ground state of the $m = 3$ film is a magnetic insulator. The Hall resistance $\rho_{xy}(B)$ could not be detected for $m = 1$ and 2 films due to their high resistance. Nevertheless, from the systematic evolution of the insulating state from $m = 3$ to $m = 1$ seen in $\rho(T)$, it is likely that the ground states of $m = 1$ and 2 are essentially the same as that of $m = 3$ film and therefore magnetic insulators. A metal to magnetic insulator transition upon changing the thickness of SrIrO$_3$ layers has been reported also in (001)-oriented superlattice of [(SrIrO$_3$)$_m$, SrTiO$_3$] [29]. In contrast to the (111) superlattice, the magnetic moment in (001) superlattices was observed only within the film plane.

The appearance of the magnetic and insulating ground states implies the importance of electron correlations in (111)-oriented CSIO superlattices. Recent theoretical calculations indeed indicated that (111) bilayer ($m = 1$) of SrIrO$_3$ perovskite is on the verge of an orbital selective topological Mott transition and that a reasonable Coulomb $U \sim 2$ eV switches a topological insulator to a trivial insulator with the help of magnetic ordering [30]. This may provide a reasonable explanation for our observations on CSIO superlattices. In this context, it is tempting to investigate the suppression of the magnetic ordering, for example, by physical pressure to switch the system back to the topological state.

In conclusion, by optimizing the size of the alkaline earth ions, superlattice structures with alternating 2$m$ ($m$ = 1, 2, and 3) layers of (111)-oriented Ca$_{0.5}$Sr$_{0.5}$IrO$_3$ perovskite and bilayer of SrTiO$_3$ were successfully stabilized on STO(111) substrates. The structural characterization by XRD and STEM indicated that the superlattice structures were well controlled on an atomic level. The ground states of the superlattice films were found to be magnetic insulators. This may imply the importance of electron correlations which were proposed to compete with the topological effect expected from the honeycomb-based lattice structure of Ir in (111) superlattices. This study has established experimentally that (111) perovskite superlattice structures can be a realistic playground for exploring the interplay of electron correlations and topological effects.

**Acknowledgements**

The authors are grateful to A. S. Gibbs for valuable comments. This work was supported by a Grant-in-Aid for Scientific Research (grant number 24224010 and 25103724) from MEXT, Japan.

# References

[1] D. G. Schlom, L.-Q. Chen, X. Pan, A. Schmehl, and M. A. Zurbuchen, J. Am. Ceram. Soc. **91**, 2429 (2008).

[2] P. Zubko, S. Gariglio, M. Gabay, P. Ghosez, and J.-M. Triscone, Ann. Rev. Condens. Matter Phys. **2**, 141 (2011).

[3] J. Chakhalian, A. J. Millis, and J. Rondinelli, Nat Mater **11**, 92 (2012).

[4] H. Y. Hwang, Y. Iwasa, M. Kawasaki, B. Keimer, N. Nagaosa, and Y. Tokura, Nat. Mater. **11**, 103 (2012).

[5] D. Xiao, W. Zhu, Y. Ran, N. Nagaosa, and S. Okamoto, Nat. Commun. **2**, 596 (2011).

[6] A. Rüegg and G. A. Fiete, Phys. Rev. B **84**, 201103 (2011).

[7] S. Okamoto, Phys. Rev. Lett. **110**, 066403 (2013).

[8] D. Doennig, W. E. Pickett, and R. Pentcheva, Phys. Rev. Lett. **111**, 126804 (2013).

[9] A. H. Castro Neto, F. Guinea, N. M. R. Peres, K. S. Novoselov, and A. K. Geim, Rev. Mod. Phys. **81**, 109 (2009).

[10] J. Chang, J.-W. Lee, and S.-K. Kim, J. Cryst. Growth **312**, 621 (2010).

[11] B. Gray, H. N. Lee, J. Liu, J. Chakhalian, and J. W. Freeland, Appl. Phys. Lett. **97**, 013105 (2010).

[12] M. Gibert, P. Zubko, R. Scherwitzl, J. Íñiguez, and J.-M. Triscone, Nat. Mater. **11**, 195 (2012).

[13] G. Herranz, F. Sánchez, N. Dix, M. Scigaj, and J. Fontcuberta, Sci. Rep. **2**, (2012).

[14] S. Middey, D. Meyers, M. Kareev, E. J. Moon, B. A. Gray, X. Liu, J. W. Freeland, and J. Chakhalian, Appl. Phys. Lett. **101**, 261602 (2012).

[15] S. Middey, D. Meyers, D. Doennig, M. Kareev, X. Liu, Y. Cao, P. J. Ryan, R. Pentcheva, J. W. Freeland, and J. Chakhalian, arXiv:1407.1570 [cond-Mat] (2014).

[16] J. Chang, Y.-S. Park, and S.-K. Kim, Appl. Phys. Lett. **92**, 152910 (2008).

[17] J. M. Longo, J. A. Kafalas, and R. J. Arnott, J. Solid State Chem. **3**, 174 (1971).

[18] A. Sumi, Y. K. Kim, N. Oshima, K. Akiyama, K. Saito, and H. Funakubo, Thin Solid Films **486**, 182 (2005).

[19] H. Gao and W. D. Nix, Ann. Rev. Mater. Sci. **29**, 173 (1999).

[20] D. J. Srolovitz, Acta Metall. **37**, 621 (1989).

[21] J. Stangl, V. Holý, and G. Bauer, Rev. Mod. Phys. **76**, 725 (2004).

[22] J.-G. Cheng, J.-S. Zhou, J. B. Goodenough, Y. Sui, Y. Ren, and M. R.

Suchomel, Phys. Rev. B **83**, 064401 (2011).

[23] S. Y. Jang, H. Kim, S. J. Moon, W. S. Choi, B. C. Jeon, J. Yu, and T. W. Noh, J. Phys.: Condens. Matter **22**, 485602 (2010).

[24] D. Hirai, J. Matsuno, and H. Takagi, arXiv:1501.01433 [cond-Mat] (2015).

[25] S. J. Moon, H. Jin, K. W. Kim, W. S. Choi, Y. S. Lee, J. Yu, G. Cao, A. Sumi, H. Funakubo, C. Bernhard, and T. W. Noh, Phys. Rev. Lett. **101**, 226402 (2008).

[26] J. Liu, J.-H. Chu, C. R. Serrao, D. Yi, J. Koralek, C. Nelson, C. Frontera, D. Kriegner, L. Horak, E. Arenholz, J. Orenstein, A. Vishwanath, X. Marti, and R. Ramesh, arXiv:1305.1732 [cond-Mat] (2013).

[27] M. A. Zeb and H.-Y. Kee, Phys. Rev. B **86**, 085149 (2012).

[28] Y. F. Nie, P. D. C. King, C. H. Kim, M. Uchida, H. I. Wei, B. D. Faeth, J. P. Ruf, J. P. C. Ruff, L. Xie, X. Pan, C. J. Fennie, D. G. Schlom, and K. M. Shen, Phys. Rev. Lett. **114**, 016401 (2015).

[29] J. Matsuno, K. Ihara, S. Yamamura, H. Wadati, K. Ishii, V. V. Shankar, H.-Y. Kee, and H. Takagi, arXiv:1401.1066 [cond-Mat] (2014).

[30] S. Okamoto, W. Zhu, Y. Nomura, R. Arita, D. Xiao, and N. Nagaosa, Phys. Rev. B **89**, 195121 (2014).